\documentclass[twoside]{ilcws08}
\usepackage[latin1]{inputenc}
\usepackage[dvips]{graphicx,epsfig,color}
\usepackage{wrapfig,rotating}
\usepackage{amssymb,amsmath,array}
\usepackage{subfigure}

\def\etal{{\it et~al.}}%
\begin{document}
\title{R\&D for Very Forward Calorimeters at the ILC Detector
} 
\author{W. Lohmann\\
on behalf of the FCAL Collaboration
\thanks{http://www-zeuthen.desy.de/ILC/fcal/}
\vspace{.3cm}\\
DESY \\
Platanenallee 6, Zeuthen, Germany
\vspace{.1cm}\\
}

\maketitle

\begin{abstract}
Special calorimeters are needed to instrument the very forward region of
an ILC detector. These devices will improve the hermeticity being
important for new particle searches.
A luminometer is foreseen to measure the rate of
low angle Bhabha scattering with a precision better than 10$^{-3}$.
A calorimeter adjacent to the beam-pipe will be hit by a large amount 
of beamstrahlung remnants. The amount and shape of these
depositions will allow a fast luminosity estimate
and the determination of beam parameters. However, the sensors
must be extremely radiation hard.
Finely segmented and very compact calorimeters will match the
requirements. Due to the high occupancy 
fast FE electronics is needed.
A possible design of the calorimeters is presented
and an overview on the R\&D status is given~\cite{url}.
\end{abstract}

\section{The Challenges}

Two calorimeters are foreseen in the very forward region of the ILC detector 
near the interaction point - LumiCal for the precise measurement
of the luminosity and BeamCal for the fast estimate of the luminosity~\cite{ieee1}. 
Both will improve the hermeticity of the detector. 
A third calorimeter, GamCal, 
about 100 m downstream of the detector, will assist beam-tuning.
Also for beam-tuning a pair monitor is foreseen, positioned just 
in front of BeamCal.

LumiCal 
will measure the luminosity using as gauge 
process Bhabha 
scattering, $e^+e^- \rightarrow e^+e^-(\gamma)$.
To match the physics benchmarks,
an 
accuracy of better than
10$^{-3}$ is needed. For the GigaZ option even an accuracy
of 10$^{-4}$ is aimed~\cite{klaus}. Hence, LumiCal is a precision device
with challenging  
requirements on the mechanics 
and position control.
All detectors in the very forward region have to tackle relatively high 
occupancies, requiring special FE electronics and data transfer equipment. 

BeamCal is positioned
just outside the beam-pipe. A large amount of low energy 
electron-positron pairs 
originating from beamstrahlung will deposit their energy
in BeamCal. These depositions, useful for a 
bunch-by-bunch luminosity estimate and the determination of 
beam parameters~\cite{grah1}, 
will lead, however, to a 
radiation dose of several MGy per year in the sensors
at lower polar angles.
Hence extremely radiation hard sensors are needed to instrument 
BeamCal. A pair monitor, consisting of a layer of pixel sensors positioned 
just in front of BeamCal, will measure the distribution of beamstrahlung pairs and
give additional information for beam parameter determination.

A small Moliere radius is of invaluable importance for both calorimeters. 
It ensures an excellent
electron veto capability for BeamCal even
at small polar angles, being essential to suppress
background in new particle searches where the signatures 
are large missing energy and momentum.
In LumiCal the precise reconstruction of electron and positron 
showers of Bhabha events
is facilitated and background processes will be rejected efficiently.
\begin{wrapfigure}{r}{0.45\columnwidth}
\centerline{\includegraphics[width=0.4\columnwidth]{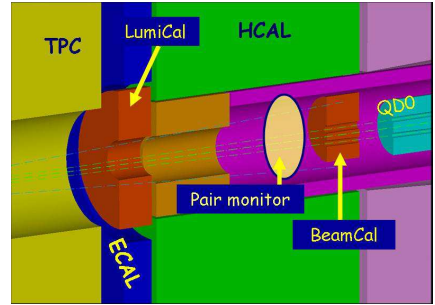}}
\caption{The very forward region of the ILD detector. LumiCal, BeamCal, Pair Monitor, and QD0 are carried by 
the support tube. 
TPC denotes the central track chamber, ECAL the elecromagnetic and 
HCAL the hadron calorimeter.}\label{fig:Forward_structure}
\end{wrapfigure}

GamCal will be positioned about 100 m downstream and 
is foreseen to measure the amount and energy distribution of beamstrahlung photons.
These are complementary informations to assist beam-tuning and determine
beam parameters at the interaction point.

\section{The Design of the Very Forward Region} 
 
A sketch of the very forward region of the ILD detector, as an example, is 
shown in 
Figure~\ref{fig:Forward_structure}. 
LumiCal and BeamCal are cylindrical electromagnetic calorimeters, 
centered around the
outgoing beam. LumiCal is positioned inside and aligned with the 
forward electromagnetic calorimeter.
BeamCal is placed just in front of the final focus quadrupole.


\subsection{LumiCal} 

\begin{wrapfigure}{r}{0.5\columnwidth}
\centerline{
\includegraphics[width=0.45\columnwidth,height=4.8cm]{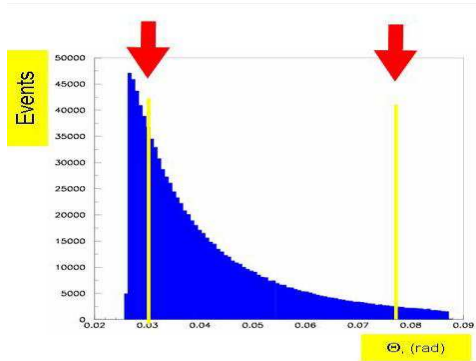}}
\caption []{\label{fig:Bhabha_theta}
The polar angle distribution of Bhabha events in the LumiCal. The arrows 
indicate the acceptance
region in the polar angle $\theta$.  
}
\end{wrapfigure} 
Monte Carlo studies have shown that a compact silicon-tungsten 
sandwich calorimeter is a proper technology
for LumiCal~\cite{Ronen}.
In the current design~\cite{woitek1}, 
LumiCal  
covers the polar angular range between 32 and 74 mrad.
The 30 layers of 
tungsten absorber disks are interspersed with silicon sensor planes. 
The FE and ADC ASICS are positioned at the outer radius in
the space between the disks.
The small Moliere radius and finely radially segmented 
silicon pad sensors ensure an efficient
selection of Bhabha events and a precise shower position measurement. 
The luminosity, $\cal{L}$, 
is obtained from 
$\cal{L} = \cal{N}/{\sigma}$,
where $\cal{N}$ is the number of Bhabha events counted in a 
certain polar angle range 
and $\sigma$ is the Bhabha scattering cross section in the 
same angular range
calculated from theory.
The most critical quantity to control when counting 
Bhabha events 
is the inner acceptance radius 
of the calorimeter, defined as the lower cut in the polar 
angle, as illustrated in   
Figure~\ref{fig:Bhabha_theta}. 
\begin{wrapfigure}[15]{r}{0.45\columnwidth}
\centerline
    {\includegraphics[width=0.4\columnwidth]{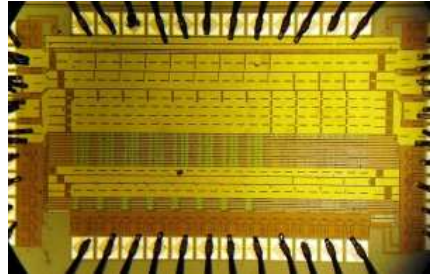}}
\caption []{\label{fig:ASICS}      
Prototypes of the FE ASIC prepared for systematic tests in the laboratory.} 
\end{wrapfigure} 
Since the angular distribution is very steep, a small 
bias in the lower polar angle measurement will
shift the measured value of the luminosity. 
From Monte Carlo studies of the given design a
tolerance of a few $\mu$m is estimated for the 
inner acceptance radius~\cite{achim1}.
Since there is bremsstrahlung radiation 
in Bhabha scattering, also cuts on the shower energy 
will be applied. The criteria to select good 
Bhabha events hence define requirements 
on the energy resolution and, more challenging, 
on the control of the energy 
scale of the calorimeter. The latter quantity must be known to about a few
per mill~\cite{ivan}.
Monte Carlo simulations are also used
to optimise the radial and azimuthal segmentation of silicon pad sensors 
for LumiCal~\cite{Ronen} to match the requirements on 
the shower position measurement performance.

A first batch of prototype sensors~\cite{woitek2}
is ordered from Hamamatsu Corp.
These sensors will be available
in spring 2009. 
At the first stage they will be studied in the laboratories. 
In a later stage, they will 
be instrumented with Front-End (FE) electronics 
for investigations in the test-beam and to prepare a calorimeter prototype
for systematic performance studies. 
FE and ADC ASICS are designed
with a shaping and conversion time less than 300 ns, being potentially able
to readout the calorimeter after each bunch crossing.
The range of sensor pad capacitance and the expected signal range
in electromagnetic showers originating from Bhabha events are taken from 
Monte-Carlo simulations
~\cite{iftacheu1}. A prototype of the FE ASIC,
manufactured in 0.35 $\mu$m
AMS technology, is shown in Figure~\ref{fig:ASICS}.
The FE ASIC can be operated in low and high amplification mode.
The high amplification
mode allows to measure the depositions of minimum ionising particles.
This option allows to use muons from the beam halo or from annihilations for the
calibration and sensor alignment studies. The low amplification mode 
will be used for the measurement of electromagnetic showers.   
Tests of these ASICS prototype are ongoing in the laboratory~\cite{FE_ASIC}. Results
on linearity, noise and cross talk measured in the laboratory
are matching the requirements for the performance derived from Monte-Carlo 
simulations.   
For 2010 multi-channel prototypes of the ASICS are planned, allowing to instrument
prototypes of sensor planes to investigate the performance of the full system in the test-beam.
For the position monitoring a dedicated laser system is under development. A
small scale prototype was designed 
and built~\cite{woitek3} and the measured performance matches the requirements.

\subsection{BeamCal}

\begin{wrapfigure}{l}{0.45\columnwidth}
\centerline{    
\includegraphics[width=0.4\columnwidth]{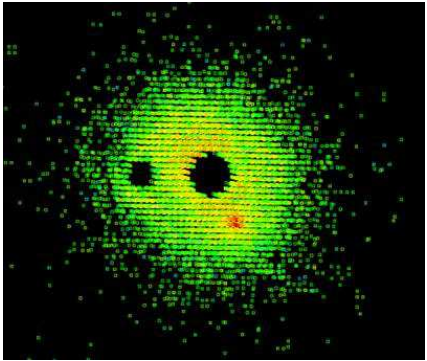}}
\caption []{\label{fig:beam_deposits}
The distribution of depositions of beamstrahlung pairs after one bunch
crossing on BeamCal. Superimposed is the deposition of
a single high energy electron (dark or red spot in the bottom part). The black holes 
correspond to the beam-pipes. 
}
\end{wrapfigure}

BeamCal is designed as a sensor-tungsten sandwich calorimeter covering 
the polar angle range between
5 and 40 mrad. 
The tungsten absorber disks will be of one radiation length thickness and 
interspersed with thin sensor layers
equipped with FE electronics positioned at the outer radius. 
In front of BeamCal an about 5 cm thick graphite block
is placed to absorb low energy back-scattered particles.
BeamCal will be hit after each bunch-crossing 
by a large amount of beamstrahlung pairs, 
as shown in Figure~\ref{fig:beam_deposits}. 
The amount, up to several TeV per bunch crossing, 
and shape of 
these deposition
allow a bunch-by-bunch 
luminosity estimate and the determination of beam parameters~\cite{grah1}.
However, depositions of  
single high energy 
electrons must be detected on top of the
wider spread beamstrahlung.
Superimposed on the pair depositions 
in Figure~\ref{fig:beam_deposits}
is the local deposition of one high energy electron, seen as 
the dark or red spot at the bottom.
Using an appropriate subtraction of the pair deposits
and a shower finding algorithm which
takes into account the longitudinal shower profile, 
the deposition of the high energy electron 
can be detected with high efficiency and modest energy resolution, 
sufficient to suppress the
background from two-photon processes in a search e.g. 
for super-symmetric tau-leptons~\cite{drugakov}. 
\begin{wrapfigure}[16]{r}{0.5\columnwidth}
\centerline{    
\includegraphics[width=0.45\columnwidth,height=4.4cm]{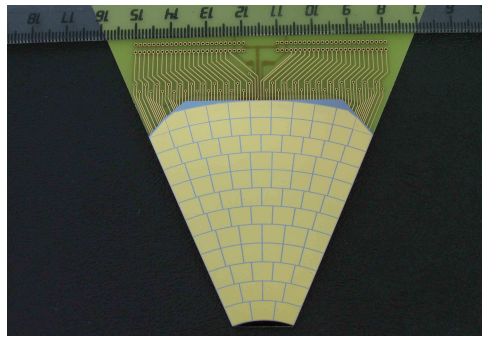}}
\caption []{\label{fig:GaAs}
A prototype of a GaAs sensor sector for BeamCal with pads of about 0.3 cm$^2$ area.   }
\end{wrapfigure} 
The challenge of BeamCal are radiation hard sensors, 
surviving up to 10 MGy of dose per year.
So far we have studied polycrystalline CVD diamond sensors of 
1 cm$^2$ size, and larger sectors of GaAs pad sensors
as shown in Figure~\ref{fig:GaAs}.
Polycrystalline CVD diamond sensors are irradiated up to 
7 MGy and are still operational~\cite{ieee2}. 
GaAs sensors are found to tolerate nearly 2 MGy~\cite{ieee3}.
Since
large area CVD diamond sensors are extremely expensive, 
they may be used only at the innermost part of
BeamCal. At larger radii GaAs sensors seem 
to be a promising option. These studies will be 
continued in future
for a better understanding of the damage mechanisms 
and possible improvements of the sensor materials.   
The FE ASIC development for BeamCal, 
including a fast analog summation for the fast beam feedback system
and an on-chip digital memory for readout in between two bunch 
trains~\cite{angel} 
is progressing, and we may expect first prototypes in 2009.

\section{The Pair Monitor}

The pair monitor consists of one layer of 
silicon pixel sensors just in front of BeamCal
to
measure the distribution of the number of beamstrahlung pairs.
Monte Carlo simulations have shown 
that the pair monitor will give essential additional information
for beam tuning. In addition, averaging over several 
bunch crossings, e.g. the 
beam sizes at the interaction point might be reconstructed with per cent precision
~\cite{sato}.
A special ASIC is developed for the pair monitor and 
prototypes are under study.
In a later stage, 
the pixel sensor and the ASIC are foreseen to be embedded in the same wafer.
The latter development will be done in SoI technology~\cite{takubo}.

\section{GamCal}

GamCal is supposed to exploit the photons 
from beamstrahlung for fast beam diagnostics.
Near the nominal luminosity the energy of beamstrahlung 
photons supplements the data from BeamCal and Pair Monitor
improving the precision of beam parameter 
measurements and reducing substantially the correlations between 
several parameters~\cite{grah1}. 
At low luminosity the amount of depositions
on BeamCal might not be sufficient 
to assist beamtuning, however GamCal will still give robust information.

To measure the beamstrahlung spectrum a small 
fraction of photons will be converted by a thin 
diamond foil or a gas-jet target about 100 m 
downstream of the interaction point. The created electrons or positrons
will be measured by an electromagnetic calorimeter. 
For the time being we have a rough design of GamCal. 
More detailed Monte Carlo studies are necessary
to fully understand the potential of GamCal for 
beam tuning and beam parameter determination.

\section{Priority R\&D Topics for FCAL }  
 
The current research work covers several fields of high priority to demonstrate that the designed devices
match the requirements from physics.
These
are:
\begin{itemize}
\item{Development of radiation hard sensors for BeamCal. 
The feasibility of BeamCal depends essentially on the availability large area radiation hard sensors.}
\item{Development of high quality sensors for LumiCal, integration of the FE electronics in a miniaturised
version and tuning of the full system to the required performance.}  
\item{Prototyping of a laser position monitoring system for LumiCal. In particular the 
control of the inner acceptance radius with $\mu$m accuracy is a challenge and must be demonstrated.}
\item{Development and prototyping of FE ASICS for BeamCal and the pair monitor. 
There are challenging requirements on the readout speed, the dynamic range, the buffering depth and 
the power dissipation. In addition, a scheme for data transfer to the back-end system has to be worked out.}
\item{Design of GamCal and estimate of its potential for a fast feedback beam-tuning system.}    
\end{itemize}

\section{Acknowledgments}

This work is supported by the Commission of the European Communities
under the 6$^{th}$ Framework Programme "Structuring the European
Research Area", contract number RII3-026126.


\begin{footnotesize}



%

\end{footnotesize}


\end{document}